\documentclass[pdftex,a4paper]{scrartcl}
\usepackage[ngerman,english]{babel}
\usepackage{amsmath}
\usepackage{graphicx}
\usepackage{amssymb}
\usepackage{hyperref}

\input epsf
\def\beq{\begin{eqnarray}}
\def\eeq{\end{eqnarray}}

\def\lsim{\mathrel{\rlap{\lower3pt\hbox{\hskip0pt$\sim$}}
    \raise1pt\hbox{$<$}}}         
\def\gsim{\mathrel{\rlap{\lower4pt\hbox{\hskip1pt$\sim$}}
    \raise1pt\hbox{$>$}}}         

\title{Aharonov-Bohm protection of black hole's baryon/skyrmion hair} 

\author{{\bf Gia Dvali$^{\textrm{a,b,c}}$ and  Alexander Gu\ss mann$^{\textrm{a}}$}}

\begin{document}

\maketitle

\centerline{\it $^{\textrm{a}}$ Arnold--Sommerfeld--Center for Theoretical Physics,}
\centerline{\it Ludwig--Maximilians--Universit\"at, 80333 M\"unchen, Germany}
\medskip
\centerline{\it $^{\textrm{b}}$ Max--Planck--Institut f\"ur Physik,
Werner--Heisenberg--Institut,}
\centerline{\it 80805 M\"unchen, Germany}
\medskip
\centerline{\it $^{\textrm{c}}$ Center for Cosmology and Particle Physics,
Department of Physics, New York University}
\centerline{\it 4 Washington Place, New York, NY 10003, USA}
\medskip

\abstract{The baryon/skyrmion correspondence implies that the baryon number is encoded into a topological surface integral. 
Under certain conditions that we clarify,  
this surface integral
can be measured by an asymptotic observer in form of an Aharonov-Bohm phase-shift in an experiment in which the skyrmion passes through
a loop of a probe string.   
  In such a setup the baryon/skyrmion number must be respected by black holes, despite the fact that it produces 
 no long-range classical field.  If initially swallowed by a black hole, the baryon number must resurface in form of a classical skyrmion hair, after the  black hole evaporates below a certain critical size.
 Needless to say, 
 the respect of the baryon number by black holes is expected to have potentially-interesting  astrophysical consequences.}

\begin{flushright}
LMU-ASC 58/16
\end{flushright}

\newpage

\section{Introduction}

 Recently \cite{us}, we have argued that the skyrmion/baryon correspondence seriously challenges the standard conclusion of the ``folk theorems" that the baryon number must be violated by black holes. We also argued that under the circumstances that will be elaborated in the present paper, this correspondence provides a ``hidden" topological protection for the baryon number. 
  The latter phenomenon is the main focus of the present paper. 
  However, before presenting our analysis let us briefly recount the  story.

    One important fact for us is the existence of 
   classical solutions of the Einstein equations describing black holes with classical skyrmion hair \cite{skyrmebh}.
    However, these solutions are only known  in the domain of parameters 
    in which the  black hole horizon $r_h$ is smaller than the characteristic skyrmion size $L$.
 As shown in \cite{us}, one can
 detect the classical skyrmion hair which is present in the regime $r_h \ll L$ through classical scattering experiments of waves scattered by the skyrmion black holes.  

 There are the two important aspects - classical and quantum - that the existence of skyrmion hair brings into the question of baryon/skyrmion number conservation in the presence of black holes.   
  
   First, as we have discussed in \cite{us}, although the very existence of a classically-observable skyrmion hair of a black hole with horizon size $r_h \ll L$  {\it a priory} does not guarantee the conservation of  baryon/skyrmion number in the presence of black holes, nevertheless this conservation gets promoted  
 into a self-consistent possibility:     
  the skyrmion/baryon charge swallowed by a large black hole 
 need not be lost, but instead can re-emerge in form of a
 black hole solution with classical skyrmion hair
after the black hole - which initially swallowed a baryon - shrinks, due to Hawking evaporation, down to the size $L$. 
     
  The reason why we cannot make a stronger statement based on purely classical considerations is that in the parameter-domain in which the black hole horizon $r_h$  largely exceeds the characteristic size of the skyrmion, $L$,  no  black hole solutions with classical skyrmion hair are known.  Hence, it is impossible 
  to classically monitor skyrmion/baryon charge when  the black hole size exceeds $L$.

 The quantum considerations bring a crucial new insight into the story.  We have argued \cite{us} that when the quantum effects 
  are taken into the account,
  even in the  domain  $r_h \gg L$ the black hole can carry a full memory of the baryon/skyrmion number it had swallowed. This memory is kept  in form of a topological  boundary surface integral that - under certain conditions - can be measured by means of an  Aharonov-Bohm phase-shift. 
  Due to its topological nature this  measurement is insensitive to a local state of the black hole-skyrmion system and allows to monitor the baryon/skyrmion charge  of a black hole also in the regime 
 $r_h \gg L$ in which the skyrmion hair is classically unobservable.   
Hence,  in this situation, the baryon/skyrmion content must be  revealed sooner or later. 
 
  If so, in what form this process should take place?  Of course, we cannot exclude at this level of the discussion a possibility that  the non-perturbative spectrum of gravity contains  objects that are more compact than QCD-skyrmions and can carry the required skyrmion/baryon charge. In such a case, the locally-observable baryon/skyrmion  charge can come out in form of these exotic creatures.   
 However, we do not see the necessity for such a scenario. 

 From our point of view it is most natural to expect that the 
 return of the ``borrowed" baryon/skyrmion  charge happens after the black hole shrinks to a size below $L$, where the known solutions with classical skyrmion hair  do exist.
 
 We must stress, however, that the end result of the black hole evolution is unimportant for the discussion of the present paper, since we focus exclusively on the topological and gauge
constraints, which must be respected for all the cases.   
 \\

\section{Setup and ingredients}

    We shall first consider skyrmions in flat space, which is enough for capturing the essence of  the topological 
    protection.  For simplicity, we restrict our analysis to the case of two quark flavors. 
    For the metric we shall use the signature $(+,-,-,-)$.

The Skyrme Lagrangian in the case of two quark flavors  is given by \cite{skyrme, Adkins:1983ya, Adkins:1983hy}
\begin{equation}
\mathcal{L}_{Skyrme} \, = \, \mathcal{L}_2 + \mathcal{L}_4 + \mathcal{L}_m \,, 
\label{lagrangian}
\end{equation}
with, 
\begin{equation}
\mathcal{L}_2= - \frac{F_\pi^2}{4}\mathrm{Tr}\left(U^+\partial_\mu U U^+ \partial^\mu U\right) 
\end{equation}
\begin{equation}
\mathcal{L}_4 = \frac{1}{32 e^2} \mathrm{Tr}\left([\partial_\mu U U^+, \partial_\nu U U^+]^2\right)
\end{equation}
\begin{equation}
\mathcal{L}_m = \frac{1}{2} m_\pi^2 F_\pi^2 \left(\mathrm{Tr} U - 2\right)
\end{equation}
where $m_\pi$ is the pion mass and $U$ is a $SU(2)$ matrix defined as,
\begin{equation}
U = e^{\frac{i}{F_\pi}\pi_a(x) \sigma_a} \,, 
\label{matrixU}
\end{equation}
with $\pi_a(x)$ the pion fields and $\sigma_a$ the Pauli matrices. $F_\pi$ is the pion decay constant  
and $e$ is the Skyrme coupling constant. 

 The above Lagrangian admits regular stable spherically-symmetric static configurations for which 
 the pion field assumes a hedgehog form, 
\begin{equation}
\frac{\pi_a}{F_\pi}= F(r) n_a \, ,
\label{hedgehog}
\end{equation}
where $n_a \equiv {x_a \over r} $ is a unit vector in radial direction and $F(r)$ is a profile function with the boundary conditions,
\begin{equation}
F(0)= B \pi,~ F(\infty)= 0 \,. 
\label{boundary}
\end{equation}
 Here $B$ is an integer, which  sets the topological charge of the soliton \cite{Adkins:1983ya}, 
\begin{equation}
B = \int d^3x J_0 \, ,
\label{Bcharge1}
\end{equation}
where $J_0$ is the time-component of the skyrme topological Chern-Simons current, 
\begin{equation}
J_\mu \, = \, - \, \frac{\epsilon_{\mu \nu \alpha \beta}}{24 \pi^2} \mathrm{Tr} \left(U^{-1}\partial^\nu U U^{-1}\partial^\alpha U U^{-1} \partial^\beta U\right)\,.
\label{current}
\end{equation}

The explicit form of the solution-profile function is not important for our purposes and the discussions can be found in \cite{Adkins:1983ya, Adkins:1983hy}.

The characteristic size and the mass of the skyrmion are given by  
\begin{equation}
L = \frac{1}{e F_\pi} \, ~{\rm and} ~ M_S = \frac{F_\pi}{e} \,,
\end{equation}
 respectively. 
\newline

 The next ingredient we shall need for our analysis is the skyrmion/baryon correspondence. 
 Although originally the skyrmions were introduced 
by  Skyrme \cite{skyrme}  as description for baryons, the correspondence is best understood in $SU(N_C)$  QCD with large number of colors after work by Witten \cite{witten}.  We shall therefore work in the regime $N_C \gg 1$.    
 
 In this case there exists a one-to-one correspondence between the  N\"other baryonic current in theory of quarks  $q_f$
 ($f \equiv$ flavor index),    
\begin{equation}
J_\mu = \frac{1}{N_C}\sum_f \bar{q}_f \gamma_\mu q_f \,, 
\label{Bcurrent}
\end{equation}
and the topological Chern-Simons current (\ref{current}) in the chiral effective theory of pions. 

 Hence, in this sense a baryon carries a topological skyrmion charge. 
  The key point now is to notice \cite{us} that this charge can be measured at infinity, despite the fact that
 neither a baryon nor the skyrmion produces a locally-observable classical field at $r \rightarrow \infty$.   
  Nevertheless, there exists a possibility of detecting the skyrmion charge in Aharonov-Bohm type experiments.

   In order to rewrite the baryon charge as a surface integral, let us first notice that the skyrmion topological current  (\ref{current})  represents a 
   Hodge-dual of the exterior derivative of a two-form which takes - when evaluated on the hedgehog ansatz (\ref{hedgehog}) with the boundary conditions (\ref{boundary}) - the following form
   \begin{equation}
   S_{\mu\nu} \equiv  - {1\over 4\pi^2} \left ( F(r) - {1\over 2} {\rm sin}(2F(r)) - \pi \right ) 
   \partial_{[\mu} {\rm cos} (\theta) \partial_{\nu]} \phi    \,.  
   \label{twoform}  
  \end{equation} 
   Here $r, \theta$ and $\phi$ are the usual spherical coordinates and $F(r)$ is 
   the  profile function, with the boundary conditions $F(0) =  \pi$ and 
   $F(\infty) = 0$.   For simplicity, we have chosen the $B=1$ case of (\ref{boundary}). 
    The choice of the constant $\pi$ in (\ref{twoform}) is uniquely dictated by 
    the requirement that the two-form  $S_{\mu\nu}$ is well-defined 
    everywhere.  This is how the information about the topological charge is encoded into 
    this two-form.  The value of the charge is insensitive to a  particular form of the function $F(r)$, but only to its 
    boundary values.   
     
   The topological charge (\ref{Bcharge1}) is then given by an integral over a boundary surface enclosing the skyrmion, which can be taken  to be a two-sphere of infinite (or sufficiently-large) radius $r$, and 
 with embedding coordinates $X^{\mu}$,   
   \begin{equation} 
     B = \int_{S_2} dX^{\mu}\wedge dX^{\nu} S_{\mu\nu}   =  1\, .  
   \label{Intboundary} 
   \end{equation}
 This gives the correct representation of the skyrmion charge (\ref{Bcharge1}) in form of the boundary surface integral.       

\section{Measuring the baryon number by Aharonov-Bohm type experiments}

 Despite the fact that there is no locally-observable classical field asymptotically, 
the topological baryonic charge is nevertheless observable via  an Aharonov-Bohm type measurement.   
    Such a measurement can be performed by coupling the two-form $S_{\mu\nu}$ to a  string, via the usual 
  coupling between the string world-sheet element and an antisymmetric two-form, 
   \begin{equation} 
     {\mathcal S}_{string}  = g \int  dX^{\mu}\wedge dX^{\nu} S_{\mu\nu} \, ,  
   \label{couplingtostring} 
   \end{equation}  
   where $X^{\mu}$ are string embedding coordinates and $g$ is a coupling constant. 
  The precise nature of the string is unimportant for 
  our present analysis. 
  For example, its role can be played by a cosmic string, a fundamental string, or any other string-like 
  object, e.g., a flux-tube of some gauge field.  
  Later we shall give  an example of a particular microscopic resolution of the probe string in form of a cosmic string. 
  However, at the moment we shall work at the level of effective theory and treat the probe string as fundamental.

      Any physical process in which the string world-volume encloses the skyrmion, will result into
      an Aharonov-Bohm effect with the phase shift given by, 
    \begin{equation}
    \Delta \Phi \, = \, 2\pi g \, , 
    \label{fractional}
    \end{equation} 
   and will be observable  provided $g$ is not an integer. 
 
\begin{figure}
\centering
\includegraphics[scale=0.4]{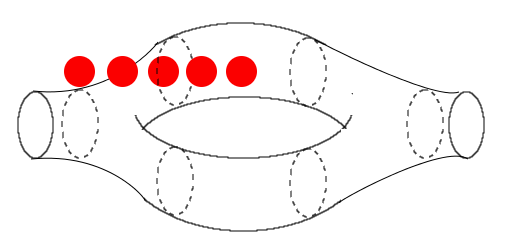}
\caption{Interference of two strings with a skyrmion (illustrated by the red circle) passing through one of them}
\label{fig:1}
\end{figure}

\begin{figure}
\centering
\includegraphics[scale=0.4]{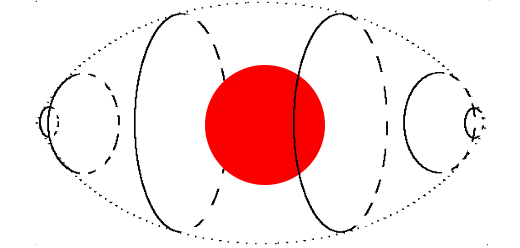}
\caption{A virtual string loop surrounding a skyrmion (illustrated by the red circle)}
\label{fig:2}
\end{figure}

    For example, we can consider an interference of the two string loops  in the process in which 
  the skyrmion passes through one of them 
  (see Fig. \ref{fig:1} for illustration).  Alternatively, we can consider a vacuum process in which 
  a virtual string loop gets created on a north pole of an imaginary two-sphere surrounding the skyrmion,  encloses the sphere  
  and collapses at the south pole (see Fig. \ref{fig:2} for illustration).  

      Thus, in such a setup an asymptotic observer can monitor the baryon charge in form of the Aharonov-Bohm skyrmion   
  hair at infinity. Now it is clear that black holes must respect this charge.  Indeed, consider a Gedankenexperiment 
 in which a skyrmion whose topological charge has been measured at infinity ends up inside a black hole of horizon 
 size $r_h \, \gg\, L$. Although there is no known solution with classical skyrmion hair outside of such a black hole, nevertheless the information 
 about the skyrmion charge is carried by the Aharonov-Bohm phase \textit{of the conserved charge (\ref{Intboundary})}, and therefore, cannot be lost.

  Hence, after evaporation such a black hole is forced to reveal its baryon/skyrmion content in some form. 
   Of course, it is most natural to assume - as it was discussed  in \cite{us} - that the reemergence of the baryon charge 
   will take place in form of the known solutions with classical skyrmion hair, once the black hole shrinks below the size $L$. 
    However, irrespectively whether nature chooses this possibility or some yet unknown way, the baryonic charge must be respected due to its topological protection by the Aharonov-Bohm phase. \\
    
\section{The secret gauge structure}

  The possibility of detecting black hole quantum numbers via an
 Aharonov-Bohm measurement has been observed previously for the following cases:  
 for two-form gauge fields both massless \cite{Bowick:1988} and massive \cite{Allen:1989kc},  
 for discrete gauge symmetries \cite{Krauss:1988zc}  and for a massive spin-2 field \cite{Dvali:2006az}. 
 The common feature in all these examples is the existence of an elementary gauge 
 field, which assumes a pure gauge form asymptotically. This looks very different  
 from the present case, since the Skyrme Lagrangian (\ref{lagrangian}) does not contain any elementary gauge degree of freedom\footnote{Another difference of the present example from the cases of \cite{Bowick:1988}, \cite{Allen:1989kc} and \cite{Dvali:2006az}
is that  in the latter cases no non-singular flat space solutions exist, and such a hair  cannot exist without a black hole, due to the need of hiding the singularity behind the black hole horizon. Correspondingly, it is not immediately clear what should be the end result of evaporation of a black hole that carries such a hair. In contrast, in the skyrmion case, the non-singular solutions do  exist in the flat space in form of skyrmions and represent a fully consistent candidate for the end-product of the black hole evaporation.
This feature is related to the fact that the object $S_{\mu\nu}$ that plays the role of the Aharonov-Bohm hair-carrying gauge field in the skyrmion case is not an elementary
field. It can assume a topologically non-trivial  asymptotically pure-gauge form 
without causing a singularity in the skyrmion solution (see below).}. 
  
 We observe that despite the fact that the Skyrme Lagrangian has no propagating 
gauge degree of freedom,  the two-form  $S_{\mu\nu}$ effectively assumes the role analogous to the one  played by a gauge potential in an ordinary Aharonov-Bohm effect.   
 
 In order to see that the analogy is deeper than what may naively seem, notice that the Skyrme topological current (\ref{current}) is invariant under the 
 gauge shift of the two-form  $S_{\mu\nu}$ by an exterior derivative of an arbitrary one-form $\zeta_{\mu}(x)$, 
 \begin{equation} 
  S_{\mu\nu} \rightarrow S_{\mu\nu} \, + \,  \partial_{\mu} \zeta_{\nu}(x) - \partial_{\nu} \zeta_{\mu}(x) \, .
  \label{shiftS}
  \end{equation}  
  Notice, that the above gauge-redundancy is itself redundant since
  $S_{\mu\nu}$ experiences the same shift under a family  of 
   $\zeta_{\nu}$-parameters related to each other by an $U(1)_{\zeta}$ gauge transformation 
   \begin{equation}
 \zeta_{\nu} \rightarrow \zeta_{\nu} + \partial_{\nu}\alpha(x) \, , 
 \label{shiftZ}
 \end{equation} 
  where $\alpha(x)$ is an arbitrary non-singular function.    
   
  If we for a moment think about $\zeta_{\mu}$ as of a vector-potential of this   
  gauge $U(1)_{\zeta}$-symmetry, then from the point of view of this $U(1)_{\zeta}$-theory the 
  skyrme configuration of the $S_{\mu\nu}$   (\ref{twoform}) for 
  $r \rightarrow \infty$ 
  coincides with the 
  abelian field-strength of Dirac's magnetic monopole:  
  $S_{\mu\nu} =\partial_{[\mu} \zeta_{\nu]}$, where the one-form $\zeta_{\nu}$ can be chosen as, 
  \begin{equation} 
  \zeta_{\nu} =  \frac{1}{4 \pi} \left( {\rm cos}(\theta) - 1\right) \partial_\nu \phi\,.
  \label{monopole} 
  \end{equation}   
 
  In this description the skyrmion charge is effectively identified with a magnetic 
  charge of $U(1)_{\zeta}$ magnetic monopole.  This way of visualizing things 
  is not creating any new conserved quantity, but it helps to understand the 
  skyrmion charge in the language of the hidden gauge structure.  
  
  Of course, as we know very well, in the case of a $U(1)_{\zeta}$ magnetic monopole
in order to deliver a non-zero magnetic charge 
  the parameter  $\zeta_{\phi}$ must become singular somewhere. 
 For example, 
  in the gauge (\ref{monopole}) this singularity manifests itself  
  in form of a
   Dirac string oriented along the
  $z<0$ semi-axes. If $\zeta_{\nu}$ were a Maxwellian  
 gauge field, the Dirac string would be rendered unobservable  
  by imposing the Dirac quantization condition  on elementary electric charges that source $\zeta_{\nu}$.

 In our case the Dirac string is automatically unobservable since $\zeta_{\nu}$ is not a gauge degree of freedom, but a gauge redundancy parameter for $S_{\mu\nu}$.
  Due to this there exist no elementary particles that are electrically charged  under the ``field" $\zeta_{\nu}$.
 Correspondingly, the  Dirac string for the $U(1)_{\zeta}$ monopole is by-default unphysical.
   
The above is the key to understanding of why - despite the fact that the skyrmion charge can be visualized as the $U(1)_{\zeta}$-monopole charge - the skyrmion solution is regular everywhere.   
The singularity in the $U(1)_{\zeta}$-monopole configuration can be attributed 
to the parameter that parameterizes the redundancy (\ref{shiftZ}) of the redundancy
(\ref{shiftS}).  This is most transparent in the
Wu-Yang
 formulation \cite{Wu:1975} in which 
in upper  ($0 < \theta < {\pi \over 2}$) and lower  ($ {\pi \over 2} < \theta < \pi $)
hemispheres the parameter $\zeta$ can be represented as 
follows 
 \begin{equation} 
  \zeta_{\nu}^{(U)} =  \frac{1}{4\pi}\left({\rm cos}(\theta) -1\right) \partial_\nu \phi, ~~ 
   \zeta_{\nu}^{(L)} =  \frac{1}{4\pi}\left({\rm cos}(\theta) + 1\right) \partial_\nu \phi, ~
\label{monopole1} 
  \end{equation} 
with $ \zeta_{\phi}^{(U)}$ and $\zeta_{\phi}^{(L)}$ at the equator differing by a single-valued gauge transformation (\ref{shiftZ}) with $\alpha = \phi$, 
   \begin{equation} 
  \left (\zeta_{\nu}^{(L)} -  \zeta_{\nu}^{(U)} \right )_{\theta = {\pi \over 2}}  \, = \, \frac{1}{2\pi} \partial_{\nu}\phi \,.
\label{junction} 
  \end{equation} 
That is, the redundancy parameter $\zeta_{\nu}$ is patched by using its own gauge-redundancy (\ref{shiftZ}).   
 
    Thus, the $U(1)_{\zeta}$-magnetic monopole configuration 
  that $S_{\mu\nu}$ assumes  for $r \rightarrow \infty$   
    corresponds - when viewed from the point of view of the gauge redundancy (\ref{shiftS}) - to a 
  {\it  locally-pure-gauge } form. 
   $S_{\mu\nu}$ assumes this form only asymptotically and smoothly departs from it 
   at the origin ($r \rightarrow 0$)  without encountering a singularity. 
  
  To summarize,  the   Aharonov-Bohm phase-shift (\ref{fractional}) effectively measures the magnetic charge of 
  an ``embedded"  $U(1)_{\zeta}$-Dirac monopole created by the vector gauge-parameter $\zeta_{\nu}$. Of course, in reality there is no classically-observable magnetic field, since 
  $\zeta_{\nu}$ is not a physical degree of freedom, but a parameter of the gauge redundancy  (\ref{shiftS}) of the skyrmion charge. 
    Thus, from the point of view of the two-form gauge symmetry (\ref{shiftS}) 
  the skyrmion configuration is asymptotically locally {\it pure-gauge}  and the Aharonov-Bohm phase-shift takes place due to the non-trivial topology, just as in case of an ordinary Aharonov-Bohm phenomenon.

\section{Manufacturing a probe string} 

 Here we shall give an example of manufacturing a probe string that is charged under 
 the skyrmion two-form $S_{\mu\nu}$.

Following \cite{Dvali:2006az},  let us show that the role of such a string can be played by a Nielsen-Olesen cosmic string of Abelian Higgs model. 
 
 The Lagrangian has a standard form, 
 \begin{equation} 
  \mathcal{L} \, = \,   |D_{\mu}H|^2 - \lambda^2( |H|^2 - v^2)^2 - F_{\mu\nu}F^{\mu\nu} \, .    
 \label{Higgs}
 \end{equation}
 Here $H$ is the complex scalar,  $D_{\mu} \, \equiv \, \partial_{\mu} -  iq A_{\mu}$ is the covariant derivative 
 and $F_{\mu\nu}  \equiv \partial_{[\mu} A_{\nu]}$ is the field strength. 
 $q$ is the gauge coupling. The system is invariant under  
 the gauge symmetry, 
 \begin{equation} 
 H \rightarrow {\rm e}^{i \omega(x)} H, ~~
  A_{\nu} \rightarrow A_{\nu} + \frac{1}{q} \partial_{\nu}\omega(x) \, , 
 \label{shiftA}
 \end{equation} 
 were $\omega$ is a gauge-transformation parameter.

  The non-zero vacuum expectation value (VEV) of the scalar, 
 $\langle |H| \rangle = v$, generates a mass gap and the resulting spectrum of the theory consists of a gauge boson 
 of mass  $qv$ and a scalar of mass $\lambda v$.  The phase of $H$ becomes a longitudinal polarization of the massive ``photon" $A_{\mu}$.   

  It is well known \cite{Nielsen}  that this theory admits the string-like soliton solutions, the so-called Nielsen-Olesen 
 vortex lines which can be viewed as quantum field theoretic versions of Abrikosov vortexes in superconductors.  These strings represent the tubes of quantized magnetic flux with the elementary unit of magnetic flux given by 
 $flux= \oint A = { 2\pi \over q}$, where the line integral is taken along a closed path 
 around the string.  The thickness of the magnetic flux-tube is of the order 
 of the Compton wavelength of the gauge field, $\sim (qv)^{-1}$.  
 
  In order for the Nielsen-Olesen string to act as a source for the skyrmion two-form, 
 we need to introduce the following coupling, 
 \begin{equation} 
 c\,  S_{\mu\nu} F_{\alpha\beta}  \epsilon^{\mu\nu\alpha\beta} \, ,  
 \label{SF} 
 \end{equation}
where $c$ is a parameter.   At the level of an effective theory, the above coupling 
is fully legitimate and is compatible with all the symmetries of the problem. 
For example, it is invariant under the gauge shift of the skyrmion 
two-form  (\ref{shiftS}).   

  If we now perform the above-described Aharonov-Bohm-type measurement of the skyrmion charge by using the loop of the cosmic string, the resulting phase-shift will
 be equal to 
  \begin{equation}
    \Delta \Phi \, = \, 2\pi {c \over q} \,.  
    \label{fractional2}
    \end{equation} 
  This is not surprising, since for the string loops larger than their thickness
 the coupling (\ref{SF}) reduces to an effective coupling (\ref{couplingtostring}) 
 with $g = {c \over q}$, 
  \begin{equation} 
  c\,  S_{\mu\nu} F_{\alpha\beta}  \epsilon^{\mu\nu\alpha\beta} \rightarrow  {c \over q}  \int  dX^{\mu}\wedge dX^{\nu} S_{\mu\nu} \,.   
   \label{couplingtoostring} 
   \end{equation}
   Thus, the existence of strings with fractional two-form charge $g$
  is determined by the ratio $c/q$. 
  
  The reason of why the coupling  (\ref{SF}) between the skyrmion two-form $S_{\mu\nu}$ and  the abelian field strength results into the observability of the skyrmion charge by boundary measurement can be understood in the following way. 
  The coupling (\ref{SF}) implies that the skyrme current sources $A_{\mu}$. 
 In this way, the skyrmion     
  becomes effectively {\it electrically} charged under the massive $A_{\mu}$ and the Aharonov-Bohm effect in which the string loop encloses the skyrmion can be equivalently understood as the ``ordinary" Aharonov-Bohm effect in which an object 
 (i.e., the skyrmion) which is electrically charged  with respect to $A_{\mu}$  goes through the loop of a solenoid  (i.e., the cosmic string) that carries the magnetic flux of $A_{\mu}$.  
 
 For the rational fractional values of $c/q$, we can interpret 
 the situation in the following way.  By introducing the coupling (\ref{SF}) the baryon/skyrmion acquires a fractional charge  under the gauge $U(1)$-symmetry.  The VEV of the Higgs is invariant  under a  discrete subgroup  $Z_N$, where $N$ is the minimal integer number for which $Nc/q$ is an integer.  Thus, the  
 skyrmion acquires a  discrete gauge hair ala Krauss and Wilczeck \cite{Krauss:1988zc}. 
 For irrational values of $c/q$, the subgroup is formally $Z_{\infty}$ and the skyrmion acquires an infinite discrete hair.

   An alternative useful interpretation of why in the presence of the coupling (\ref{SF}) the skyrmion acquires a fractional  charge under the Higgsed $U(1)$-symmetry  (\ref{shiftA}), is in terms of 
   Witten's  effect \cite{Witten:1979ey}. 
   As explained above,  the asymptotic configuration of $S_{\mu\nu}$ is given by the  field strength of a magnetic monopole of $\zeta_{\nu}$. Thus, asymptotically the coupling (\ref{SF}) 
 reduces to 
   a dual coupling between the  field-strengths of the $A_{\mu}$     
and $\zeta_{\mu}$ vectors, 
 \begin{equation} 
 c\,  S_{\mu\nu} F_{\alpha\beta}  \epsilon^{\mu\nu\alpha\beta} \, 
 \rightarrow 
 c\,  F^{(\zeta)}_{\mu\nu} F_{\alpha\beta}  \epsilon^{\mu\nu\alpha\beta} \,,  
 \label{SF1} 
 \end{equation}
where $F^{(\zeta)}_{\mu\nu} \equiv \partial_{[\mu}\zeta_{\nu]}$. 
 As a result, the $U(1)_{\zeta}$-monopole acquires an electric charge 
 $c/q$ under the gauge symmetry (\ref{shiftA}) through 
 Witten-type  effect.
 
 Notice, for our purposes of external monitoring of the skyrmion 
hair, the coupling $c$ in (\ref{SF}) can be taken arbitrarily small.

\section{Concluding remarks}

We conclude with some remarks in order to point out possible generalizations of our analysis, to emphasize the relevance of the effective-theory framework we are working in, to point out the importance of the large-$N_C$ approximation and to mention possible further implications of our results.
\newline

First, for simplicity we restricted our analysis to the case of a single  skyrmion with topological charge $B=1$. One can generalize our arguments both for the case with more $B=1$ skyrmions as well as for the case of a single skyrmion with $B > 1$. For a general value $B$ of the skyrmion charge the two-form (\ref{twoform}) takes the form
   \begin{equation}
   S_{\mu\nu} \equiv  - {1\over 4\pi^2} \left ( F(r) - {1\over 2} {\rm sin}(2F(r)) - B\pi \right ) 
   \partial_{[\mu} {\rm cos} (\theta) \partial_{\nu]} \phi    \,.  
  \end{equation}
The Aharonov-Bohm phase shift is then $\Delta \Phi \, = \, 2\pi g B$. Thus, if  $g$ is a rational number,  one can always 
  choose the value of $B$ for which $gB = n$, with $n$ an integer,  and the resulting phase shift $\Delta \Phi \, = \, 2\pi n$ is unobservable. This can never be  achieved if $g$ is irrational.  
  In our effective field theory treatment, since the string is a spectator and $g$ can be taken to be arbitrary, the phase shift can be made observable for any value of the skyrmion charge.
\newline

Second, we want to emphasize that both our arguments in favor of baryon/skyrmion number conservation by semi-classical black holes \cite{us} and the discussion about the measurement of the baryon/skyrmion number of a black hole by the Aharonov-Bohm type experiments which we described in this paper are based on \textit{a chiral low-energy effective theory of pions}. Our arguments do neither imply that baryon number is necessarily conserved at the level of a fundamental theory operating at higher energy scales, nor do they imply that such a high-energy theory necessarily allows to write down a coupling of the form (\ref{couplingtostring}) with $g$ non-integer.

In fact, there are well-known examples in which operators violating baryon number are generated by high-energy physics (for example this is the case in grand-unified theories like $SU(5)$).

If the effective chiral theory with skyrmions is at higher energies embedded in such a baryon-number violating theory, by consistency, the same theory must provide 
a super-selection rule that forbides the coupling  
of the form (\ref{couplingtostring}) with $g$ non-integer. In such a case no baryon number can be measured in the Aharonov-Bohm type experiments described above, because either 
the coupling (\ref{couplingtostring}) will not exist or 
the generated phase-shift would always be an integer multiple of $2 \pi$.\footnote{In cases when the high 
energy theory violates baryon number, this violation 
also penetrates in the low energy theory of pions.  
 This is the case for example \cite{Callan:1983nx} in a system with $SU(5)$-magnetic monopole,  which catalyses baryon number violation 
\cite{Callan:1982ac}. As shown in \cite{Callan:1983nx},
 in such a case, the violation of baryon number by the high energy theory translates in the low energy theory of pions 
as non-existence of a conserved gauge invariant topological current which would be  well-defined everywhere. Therefore, in such a case, it is not possible to write down a two-form of the type (\ref{twoform}) which would be well-defined everywhere and thus no corresponding conserved charge can be monitored at infinity.}
\newline

Third, we work with a large number of colors $N_C \gg 1$. In this regime the baryon/skyrmion correspondence is best understood.  For smaller values of $N_C$ it was expected that the correspondence between baryons and skyrmions still holds, but that skyrmions which correspond to the baryons require a quantum-corrected description \cite{witten}. These quantum corrections are expected to appear as finite-$N_C$ corrections.
In pure $SU(N_C)$-QCD the baryon number is an anomaly-free symmetry and must be conserved in full quantum theory. 
Thus, as long as the baryon/skyrmion correspondence holds, the  
finite-$N_C$ corrections (either $1/N_C$ or $exp(-N_c)$ type) are not expected to destabilize the skyrmion. 
 In pure QCD the topological stability of the skyrmion could be jeopardized  only by the effects that abolish the baryon/skyrmion correspondence.  The instability scale will then be determined by the strength of these effects.
\newline

 Further, we would like to comment that the skyrmion/baryon hair 
 discussed in the present work is different from the 
 quantum baryonic hair of a black hole suggested earlier in 
 \cite{GiaCesar} in the context of a specific microscopic theory.
 The two ideas are fully consistent and may very well turn out to  be complementary,  
  but at no point in the present 
 work we have made any assumptions about the quantum
 properties of a black hole.  
 The beauty of the skyrmion/baryon hair is that  it does not rely    
 on any particular microscopic picture of a black hole and  employes solely the power of topology and gauge redundancy, which must be respected by any consistent microscopic theory.      
\newline

Finally, it would be important to study further the possible astrophysical implications of the topological protection of baryon number against black holes.

\section*{Acknowledgments}
The work of G. D. was supported by Humboldt Foundation under Alexander von Humboldt Professorship, by European Commission under ERC Advanced Grant 339169 ``Selfcompletion", by DFG SFB/TRR 33 ``The Dark Universe" and by the DFG cluster of excellence EXC 153 ``Origin and Structure of the Universe". The work of A. G. was supported by the DFG cluster of excellence EXC 153 ``Origin and Structure of the Universe" and by Humboldt Foundation.


\begin{thebibliography}{1}

\bibitem{us} 
  G.~Dvali and A.~Gu{\ss}mann,
  ``Skyrmion Black Hole Hair: Conservation of Baryon Number by Black Holes and Observable Manifestations,''
  Nucl.\ Phys.\ B {\bf 913}, 1001 (2016)
  doi:10.1016/j.nuclphysb.2016.10.017
  [arXiv:1605.00543 [hep-th]].


\bibitem{skyrmebh}
  H.~Luckock and I.~Moss,
  ``Black Holes Have Skyrmion Hair,''
  Phys.\ Lett.\ B {\bf 176} (1986) 341.
  doi:10.1016/0370-2693(86)90175-9

  S.~Droz, M.~Heusler and N.~Straumann,
  ``New black hole solutions with hair,''
  Phys.\ Lett.\ B {\bf 268} (1991) 371.
  doi:10.1016/0370-2693(91)91592-J

  P.~Bizon and T.~Chmaj,
  ``Gravitating skyrmions,''
  Phys.\ Lett.\ B {\bf 297} (1992) 55.
  doi:10.1016/0370-2693(92)91069-L

  M.~Heusler, S.~Droz and N.~Straumann,
  ``Linear stability of Einstein Skyrme black holes,''
  Phys.\ Lett.\ B {\bf 285} (1992) 21.
  doi:10.1016/0370-2693(92)91294-J

  T.~Tamaki, K.~i.~Maeda and T.~Torii,
  ``Internal structure of Skyrme black hole,''
  Phys.\ Rev.\ D {\bf 64} (2001) 084019
  doi:10.1103/PhysRevD.64.084019
  [gr-qc/0106008].

  N.~Shiiki and N.~Sawado,
  ``Black holes with skyrme hair,''
  gr-qc/0501025.

  C.~Adam, O.~Kichakova, Y.~Shnir and A.~Wereszczynski,
  ``Hairy black holes in the general Skyrme model,''
  Phys.\ Rev.\ D {\bf 94} (2016) no.2,  024060
  doi:10.1103/PhysRevD.94.024060
  [arXiv:1605.07625 [hep-th]].

  S.~B.~Gudnason, M.~Nitta and N.~Sawado,
  ``Black hole Skyrmion in a generalized Skyrme model,''
  JHEP {\bf 1609} (2016) 055
  doi:10.1007/JHEP09(2016)055
  [arXiv:1605.07954 [hep-th]].

N.~Sawado, N.~Shiiki, K.~i.~Maeda and T.~Torii,
  ``Regular and black hole Skyrmions with axisymmetry,''
  Gen.\ Rel.\ Grav.\  {\bf 36} (2004) 1361
  doi:10.1023/B:GERG.0000022392.89396.4b
  [gr-qc/0401020].

  N.~Shiiki and N.~Sawado,
  ``Black hole skyrmions with negative cosmological constant,''
  Phys.\ Rev.\ D {\bf 71} (2005) 104031
  doi:10.1103/PhysRevD.71.104031
  [gr-qc/0502107].

N.~Shiiki and N.~Sawado,
  ``Regular and black hole solutions in the Einstein-Skyrme theory with negative cosmological constant,''
  Class.\ Quant.\ Grav.\  {\bf 22} (2005) 3561
  doi:10.1088/0264-9381/22/17/015
  [gr-qc/0503123].

  Y.~Brihaye and T.~Delsate,
  ``Skyrmion and Skyrme-black holes in de Sitter spacetime,''
  Mod.\ Phys.\ Lett.\ A {\bf 21} (2006) 2043
  doi:10.1142/S0217732306021426
  [hep-th/0512339].

A.~B.~Nielsen,
  ``Skyrme Black Holes in the Isolated Horizons Formalism,''
  Phys.\ Rev.\ D {\bf 74} (2006) 044038
  doi:10.1103/PhysRevD.74.044038
  [gr-qc/0603127].

\bibitem{skyrme}
  T.~H.~R.~Skyrme,
  ``A Nonlinear field theory,''
  Proc.\ Roy.\ Soc.\ Lond.\ A {\bf 260} (1961) 127.
  doi:10.1098/rspa.1961.0018


  T.~H.~R.~Skyrme,
  ``A Unified Field Theory of Mesons and Baryons,''
  Nucl.\ Phys.\  {\bf 31} (1962) 556.
  doi:10.1016/0029-5582(62)90775-7

\bibitem{Adkins:1983ya}
  G.~S.~Adkins, C.~R.~Nappi and E.~Witten,
  ``Static Properties of Nucleons in the Skyrme Model,''
  Nucl.\ Phys.\ B {\bf 228} (1983) 552.
  doi:10.1016/0550-3213(83)90559-X

\bibitem{Adkins:1983hy}
  G.~S.~Adkins and C.~R.~Nappi,
  ``The Skyrme Model with Pion Masses,''
  Nucl.\ Phys.\ B {\bf 233} (1984) 109.
  doi:10.1016/0550-3213(84)90172-X

\bibitem{witten}
  E.~Witten,
  ``Global Aspects of Current Algebra,''
  Nucl.\ Phys.\ B {\bf 223} (1983) 422.
  doi:10.1016/0550-3213(83)90063-9

  E.~Witten,
  ``Current Algebra, Baryons, and Quark Confinement,''
  Nucl.\ Phys.\ B {\bf 223} (1983) 433.
  doi:10.1016/0550-3213(83)90064-0

  E.~Witten,
  ``Skyrmions And Qcd,''
  In *Treiman, S.b. ( Ed.) Et Al.: Current Algebra and Anomalies*, 529-537 and Preprint - WITTEN, E. (84,REC.OCT.) 7p

\bibitem{Bowick:1988}
  M.~J.~Bowick, S.~B.~Giddings, J.~A.~Harvey, G.~T.~Horowitz and A.~Strominger,
  ``Axionic Black Holes and a Bohm-Aharonov Effect for Strings,''
  Phys.\ Rev.\ Lett.\  {\bf 61} (1988) 2823.
  doi:10.1103/PhysRevLett.61.2823

\bibitem{Allen:1989kc}
  T.~J.~Allen, M.~J.~Bowick and A.~Lahiri,
  ``Axionic Black Holes From Massive Axions,''
  Phys.\ Lett.\ B {\bf 237} (1990) 47.
  doi:10.1016/0370-2693(90)90459-J

\bibitem{Dvali:2006az}
  G.~Dvali,
  ``Black holes with quantum massive spin-2 hair,''
  Phys.\ Rev.\ D {\bf 74} (2006) 044013
  doi:10.1103/PhysRevD.74.044013
  [hep-th/0605295].

\bibitem{Krauss:1988zc}
  L.~M.~Krauss and F.~Wilczek,
  ``Discrete Gauge Symmetry in Continuum Theories,''
  Phys.\ Rev.\ Lett.\  {\bf 62} (1989) 1221.
  doi:10.1103/PhysRevLett.62.1221

J.~Preskill and L.~M.~Krauss,
  ``Local Discrete Symmetry and Quantum Mechanical Hair,''
  Nucl.\ Phys.\ B {\bf 341} (1990) 50.
  doi:10.1016/0550-3213(90)90262-C

S.~R.~Coleman, J.~Preskill and F.~Wilczek,
  ``Quantum hair on black holes,''
  Nucl.\ Phys.\ B {\bf 378} (1992) 175
  doi:10.1016/0550-3213(92)90008-Y
  [hep-th/9201059].

\bibitem{Wu:1975}
  T.~T.~Wu and C.~N.~Yang,
  ``Concept of Nonintegrable Phase Factors and Global Formulation of Gauge Fields,''
  Phys.\ Rev.\ D {\bf 12} (1975) 3845.
  doi:10.1103/PhysRevD.12.3845

 T.~T.~Wu and C.~N.~Yang,
  ``Dirac Monopole Without Strings: Monopole Harmonics,''
  Nucl.\ Phys.\ B {\bf 107} (1976) 365.
  doi:10.1016/0550-3213(76)90143-7

\bibitem{Nielsen}
H.~B.~Nielsen and P.~Olesen,
  ``Vortex Line Models for Dual Strings,''
  Nucl.\ Phys.\ B {\bf 61} (1973) 45.
  doi:10.1016/0550-3213(73)90350-7

\bibitem{Witten:1979ey}
  E.~Witten,
  ``Dyons of Charge e theta/2 pi,''
  Phys.\ Lett.\  {\bf 86B} (1979) 283.
  doi:10.1016/0370-2693(79)90838-4

\bibitem{Callan:1983nx}
  C.~G.~Callan, Jr. and E.~Witten,
  ``Monopole Catalysis of Skyrmion Decay,''
  Nucl.\ Phys.\ B {\bf 239} (1984) 161.
  doi:10.1016/0550-3213(84)90088-9

\bibitem{Callan:1982ac}

V.~A.~Rubakov,
  ``Superheavy Magnetic Monopoles and Proton Decay,''
  JETP Lett.\  {\bf 33} (1981) 644
   [Pisma Zh.\ Eksp.\ Teor.\ Fiz.\  {\bf 33} (1981) 658].

 
 C.~G.~Callan, Jr.,
  ``Monopole Catalysis of Baryon Decay,''
  Nucl.\ Phys.\ B {\bf 212} (1983) 391.
  doi:10.1016/0550-3213(83)90677-6

\bibitem{GiaCesar}
G.~Dvali and C.~Gomez,
  ``Black Hole's 1/N Hair,''
  Phys.\ Lett.\ B {\bf 719} (2013) 419.
  doi:10.1016/j.physletb.2013.01.020
  [arXiv:1203.6575 [hep-th]].



\end{thebibliography}
\end{document}